\documentclass[reprint, amsmath, amssymb, aip, cha]{revtex4-2}

\usepackage{graphicx} 
\usepackage{array}
\usepackage{dcolumn} 
\usepackage{bm}     
\usepackage{caption}  
\usepackage{hyperref}   
\usepackage[all]{hypcap} 
\usepackage{xcolor}   
\usepackage{float}
\usepackage{placeins}

\hypersetup{
    colorlinks=true,
    linkcolor=blue,
    filecolor=magenta,      
    urlcolor=blue,
    citecolor=blue,
}

\begin{document}

\title{Breakdown of Adiabatic Scaling and Noise-induced Functional Synchronization in Deeply Quiescent Excitable Systems}

\author{Yefan Wu}
\email{wuyefan718@gmail.com}
\email{yewu0336@uni.sydney.edu.au} 
\affiliation{Department of Mathematics and Statistics, The University of Sydney, NSW 2006, Australia}

\begin{abstract}
Coherence resonance (CR) characterizes noise-induced regularity in excitable systems, yet its evaluation in quiescent biological media is often obscured by flattened energy landscapes and complex nonlinear dynamics. In this study, we investigate the stochastic dynamics of a 3D Sherman-Rinzel-Keizer (SRK) model driven by multiplicative Feller noise. We show that traditional extremal evaluations of coherence resonance encounter a bathtub effect, characterized by a broad resonance valley, which can lead to statistical inaccuracies. To address this, we propose a logarithmic centroid extraction method, which filters out stochastic jitter and recovers the underlying adiabatic Kramers scaling with high linearity ($R^2 > 0.95$). Furthermore, we identify the physical boundary where this adiabatic approximation breaks down under the strong-noise limit. Extending our analysis to gap-junction coupled systems, we observe a noise-induced transition from sub-threshold physiological shivering (characterized by statistical correlation but negligible functional output) to macroscopic functional synchronization. Our results provide a mathematical framework for extracting optimal noise intensities in broad energy valleys and offer insights into how quiescent biological systems utilize stochastic fluctuations for functional recovery.
\end{abstract}

\maketitle

\begin{quotation}
\textbf{Microscopic fluctuations are typically viewed as a nuisance in biological signal processing; however, in nonlinear systems, they can facilitate functional order. In quiescent excitable systems, such as pancreatic $\beta$-cell clusters under metabolic stress, traditional adiabatic approximations often fail due to scale mismatches in energy landscapes. Here, we show that simple extremal evaluations of coherence resonance are complicated by a flat ``bathtub effect''. By introducing a centroid extraction method, we identify the specific boundary where Kramers scaling breaks down. Furthermore, we demonstrate a transition from sub-threshold physiological shivering to noise-induced functional synchronization, offering a theoretical perspective on how biological systems leverage noise for the restoration of functional activity.}
\end{quotation}

\section{Introduction}

The constructive role of stochastic fluctuations in nonlinear dynamical systems has advanced our understanding of signal processing and self-organization in complex environments. At the heart of this paradigm is coherence resonance (CR), a counter-intuitive phenomenon wherein an optimal level of intermediate noise maximizes the temporal regularity of an excitable system without any external periodic forcing \cite{pikovsky1997coherence, lindner2004effects}. By transforming random environmental perturbations into ordered limit-cycle excursions, CR exemplifies how dissipative systems exploit noise to sustain macroscopic rhythms \cite{deville2005transitions, korneev2024levy}. 

Such noise-induced dynamics are relevant in biological excitable networks, where functional resilience relies on the robust synchronization of heterogeneous units \cite{arenas2008synchronization, boccaletti2006complex, ryabov2025nonlocal}. A representative example is the pancreatic $\beta$-cell network, whose pulsatile bursting is governed by complex fast-slow timescale separations \cite{sherman1988emergence, bertram2000phantom, bertram2004calcium}. Under metabolic stress or inhibition, these cells are pushed into quiescent states characterized by high energy barriers. While recent studies have explored macroscopic bursting synchronization mediated by gap junctions and paracrine signaling \cite{sterk2024network, pourhosseinzadeh2025heterogeneity}, the physical boundary of how an inhibited biological network is functionally restored by boundary-sensitive multiplicative noise remains to be fully elucidated. 

Understanding the robust restoration of pulsatile rhythms in such deeply quiescent states requires analyzing the rare-event transitions over these metabolic barriers. Classically, the stochastic escape over a potential barrier is governed by Kramers' reaction-rate theory \cite{kramers1940brownian, hanggi1990reaction}, which relies on a quasi-stationary adiabatic approximation. However, this adiabatic elimination can break down in perturbed, high-dimensional excitable systems, particularly those involving multi-scale interactions \cite{zhou2012quasi, wechselberger2020geometric, borner2024saddle}. Modern investigations into quasipotentials for coupled escape problems indicate that noise alters the transition pathways, steering trajectories away from classical deterministic manifolds \cite{ashwin2023quasipotentials}. Furthermore, we show that evaluating the optimal noise intensity in these valleys is complicated by a ``bathtub effect'': the macroscopic resonance valley becomes broadened and flattened. Consequently, traditional extremal search algorithms can fail, misinterpreting discrete stochastic jitter as physical shifts and obfuscating the underlying scalings \cite{xu2025universal}.

In complex nonlinear networks, extracting optimal noise intensities and reconstructing the quasipotential landscape typically require computationally intensive global sampling frameworks, such as transition-path theory \cite{e2010transition} or umbrella sampling \cite{kastner2011umbrella}. To circumvent these costs, particularly when numerically integrating multiplicative Feller noise near physical boundaries \cite{feller1951two, horsthemke1984noise, lord2010comparison}, we propose a logarithmic centroid extraction method. By treating the flat resonance valley as a geometric entity, this approach filters out local stochastic spurious minima and captures the system's macroscopic center of mass in the parameter space.

In this paper, we employ a 3D Sherman-Rinzel-Keizer (SRK) model driven by domain-bounded Feller noise to investigate the breakdown of adiabatic scaling and the emergence of network synchronization. By applying the centroid method, we recover the underlying linear Kramers scaling ($R^2 > 0.95$) and identify the threshold where the strong-noise limit triggers an adiabatic breakdown \cite{berglund2006noise}. Furthermore, expanding our analysis to a gap-junction coupled network, we demonstrate a noise-induced transition: from sub-threshold shivering to macroscopic functional synchronization \cite{pecora1998master, satin2015pulsatile, jalan2019explosive}. 

The remainder of this paper is organized as follows. In Sec. II, we introduce the 3D SRK model with Feller noise and establish the theoretical framework including Fokker-Planck formalism, Kramers scaling, and transverse stability. In Sec. III, we present the structural mismatch caused by the ``bathtub effect'' and demonstrate the recovery and breakdown of Kramers scaling. In Sec. III.D, we analyze the noise-induced functional synchronization in the coupled network. Finally, discussions and conclusions are provided in Sec. IV and Sec. V, respectively.

\section{Model and Methods}

\subsection{The 3D SRK Model with Bounded Multiplicative Noise}
To investigate the breakdown of adiabatic scaling in quiescent biological networks, we employ the well-established 3D Sherman-Rinzel-Keizer (SRK) model \cite{bertram2000phantom, bertram2004calcium}, which characterizes the essential fast-slow relaxation oscillations in pancreatic $\beta$-cells. The deterministic backbone relies on a distinct timescale separation: the fast subsystem governs the rapid action potential spiking, while the slow subsystem regulates the gradual metabolic recovery \cite{kuehn2015multiple, bertram2017multi}. 

Specifically, the membrane potential $V$ and the fast potassium channel activation gating $n$ constitute the fast subsystem, whereas the ultra-slow gating variable $z$ represents the fraction of open ATP-sensitive potassium channels (or equivalently, the slow metabolic sensing dynamics). In realistic microdomains, the finite number of these slow channels subjects the system to intrinsic metabolic fluctuations \cite{fox1994emergent, schmandt2012stochastic}. To systematically emulate this, we introduce a state-dependent multiplicative noise into the slow recovery variable. The stochastic differential equations (SDEs) are interpreted in the It\^o sense. This choice is physically appropriate here because the noise term models stochastic gating of a finite population of slow ion channels, where each noise increment acts independently of the current gating state at the onset of each infinitesimal interval. Moreover, the It\^o interpretation is standard for Wright-Fisher and Feller-type diffusion processes in population genetics and biophysics, and it allows the direct construction of the conventional Fokker-Planck equation used in Section~II.B. The SDEs are formulated as:
\begin{align}
    C_m dV &= - \left[ I_{\text{Ca}}(V) + I_{\text{K}}(V, n) + I_{\text{S}}(V, z) \right] dt, \label{eq:dV} \\
    dn &= \frac{n_{\infty}(V) - n}{\tau_n} dt, \label{eq:dn} \\
    dz &= \frac{z_{\infty}(V) - z}{\tau_z} dt + \sigma \sqrt{z(1-z)} \, dW_t, \label{eq:sde_z}
\end{align}

where the ionic currents are defined by Ohm's law: $I_{\text{Ca}} = g_{\text{Ca}} m_{\infty}(V) (V - V_{\text{Ca}})$, $I_{\text{K}} = g_{\text{K}} n (V - V_{\text{K}})$, and $I_{\text{S}} = g_{\text{S}} z (V - V_{\text{K}})$. The steady-state gating functions take the Boltzmann form $x_{\infty}(V) = [1 + \exp((V_x - V)/s_x)]^{-1}$ for $x \in \{m, n, z\}$. In particular, the slow-variable activation function is
\[
z_{\infty}(V) = \left[1 + \exp\left(\frac{V_z - V}{s_z}\right)\right]^{-1},
\]
with $V_z=-45$ mV and $s_z=10$ mV, so that $0<z_{\infty}(V)<1$ for finite $V$. The model parameters are listed in Table~\ref{tab:params}.

\begin{table}[h]
\caption{3D SRK model parameters. The slow conductance $g_S$ serves as the metabolic-stress control parameter and is varied across simulations.}
\label{tab:params}
\begin{tabular}{llcl}
\hline\hline
Symbol & Description & Value & Unit \\
\hline
$C_m$       & Membrane capacitance        & 20       & pF   \\
$g_{\text{Ca}}$ & Maximal Ca$^{2+}$ conductance & 3.6   & nS   \\
$V_{\text{Ca}}$ & Ca$^{2+}$ reversal potential  & 25    & mV   \\
$g_{\text{K}}$  & Maximal K$^+$ conductance     & 10.0  & nS   \\
$V_{\text{K}}$  & K$^+$ reversal potential      & $-75$  & mV   \\
$\tau_n$    & Fast gating time constant    & 20      & ms   \\
$\tau_z$    & Slow gating time constant    & 20\,000 & ms   \\
$V_m$       & Half-activation of $m_\infty$& $-20$   & mV   \\
$s_m$       & Slope of $m_\infty$          & 12      & mV   \\
$V_n$       & Half-activation of $n_\infty$& $-16$   & mV   \\
$s_n$       & Slope of $n_\infty$          & 5.6     & mV   \\
$V_z$       & Half-activation of $z_\infty$& $-45$   & mV   \\
$s_z$       & Slope of $z_\infty$          & 10      & mV   \\
$g_S$       & Slow conductance (control)   & $2.5$--$4.5$ & nS \\
\hline\hline
\end{tabular}
\end{table}

Here, $dW_t$ denotes the standard Wiener process increment, and $\sigma$ acts as the effective noise intensity controlling the fluctuation variance. The multiplicative diffusion amplitude $\sqrt{z(1-z)}$, characteristic of Wright--Fisher or Feller-type processes \cite{feller1951two,horsthemke1984noise}, vanishes at both physical boundaries $z=0$ and $z=1$. Related bounded-noise formulations have also been studied in excitable biological networks \cite{yu2022effects}. Moreover, because $0<z_{\infty}(V)<1$, the deterministic drift of the slow variable, $(z_{\infty}(V)-z)/\tau_z$, is positive at $z=0$ and negative at $z=1$. Thus, in the continuous-time SDE, the stochastic amplitude vanishes at the two boundaries while the deterministic drift points inward toward the physical interval. The interval $z\in[0,1]$ therefore provides the natural physical state space for the slow gating variable.

While exact microscopic channel fluctuations are governed by discrete Markovian kinetics \cite{fox1994emergent}, the Feller-type diffusion provides a macroscopic continuum approximation that captures the state-dependent suppression of fluctuation variance near the physiological boundaries. This approximation preserves the essential bounded, state-dependent structure of the underlying channel fluctuations while avoiding the computational overhead of an explicit discrete-state simulation.

\begin{figure}[htbp]
    \centering
    \includegraphics[width=0.95\columnwidth]{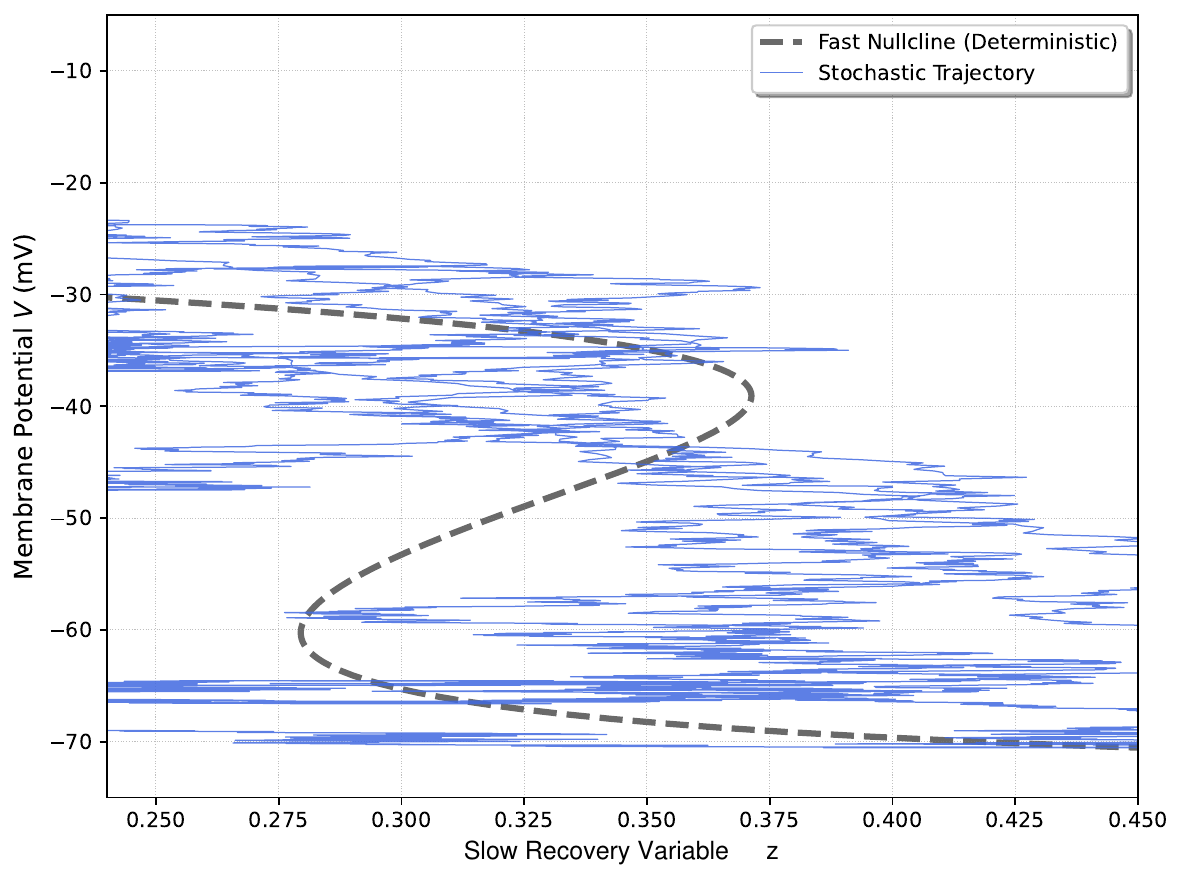}
    \caption{Projection of the stochastic 3D SRK dynamics onto the $(V, z)$ phase plane. The dashed curve shows the deterministic fast subsystem nullcline ($dV/dt = 0$), which forms a characteristic Z-shaped critical manifold. The lower and upper branches correspond to the hyperpolarized quiescent state and the depolarized active state, respectively. A representative noisy trajectory (solid line, $\sigma = 0.05$) evolves near the critical manifold and exhibits noise-induced transitions between the two dynamical regimes.}
    \label{fig:phase_plane}
\end{figure}

To visualize the geometrical foundation of the system's dynamics, we project the stochastic trajectories onto the fast-slow phase plane (Fig.~\ref{fig:phase_plane}). The deterministic fast subsystem nullcline ($dV/dt = 0$, dashed line) forms a characteristic Z-shaped manifold. The lower and upper branches represent the hyperpolarized quiescent state and the depolarized active state, respectively. Under stochastic perturbations, the trajectory evolves near the critical manifold and undergoes noise-induced transitions between the lower quiescent branch and the upper active branch, with stochastic forcing entering through the slow recovery variable $z$. This robust timescale separation forms the topological prerequisite for our subsequent analysis of Kramers scaling.

\subsection{Fokker-Planck Formalism and Feller Boundary Classification}
To justify the physical validity of the bounded multiplicative noise in Eq.~\eqref{eq:sde_z}, we invoke the Fokker-Planck equation (FPE) governing the transition probability density $P(z,t)$ \cite{freidlin2012random}. We apply an adiabatic reduction in which the fast variables $(V, n)$, which carry no direct noise term in our model, relax instantaneously to their quasi-steady state $\mathbf{x}_{ss}(z)$. We emphasize that this differs from the classical adiabatic elimination in statistical physics \cite{haken1977synergetics, gardiner2009stochastic}, where the eliminated fast variable typically carries noise (e.g., momentum in Brownian motion). Related fast-variable elimination procedures have recently been developed for stochastic systems with small inertia, including active Brownian particles with nonlinear dissipation, where elimination of the fast velocity variable yields an effective one-dimensional description \cite{goldobin2025kinetic}. Unlike the classical framework in which noise acts directly on the eliminated variable, the fast variables $(V, n)$ in our model carry no direct stochastic forcing and, under the time-scale separation approximation, are slaved to the deterministic quasi-steady manifold $\mathbf{x}_{ss}(z)$. Stochastic forcing enters only through the slow recovery variable, so that the reduced dynamics evolve along the one-dimensional slow manifold, yielding the effective 1D FPE:
\begin{equation}
    \frac{\partial P(z,t)}{\partial t} = -\frac{\partial}{\partial z} \left[ A(z) P(z,t) \right] + \frac{1}{2} \frac{\partial^2}{\partial z^2} \left[ B(z) P(z,t) \right],
    \label{eq:fpe}
\end{equation}
where the drift coefficient is $A(z) = G(\mathbf{x}_{ss}(z), z)$ and the state-dependent diffusion coefficient is $B(z) = \sigma^2 z(1-z)$. Here, $G(\mathbf{x}, z) = [z_\infty(V) - z] / \tau_z$ is the deterministic drift of the slow variable from Eq.~\eqref{eq:sde_z}, evaluated with the fast variables at their quasi-steady state $\mathbf{x}_{ss}(z) = (V_{ss}(z), n_{ss}(z))$. This reduction to a single FPE in $z$ alone is valid precisely because the fast variables $(V, n)$ carry no noise term and relax rapidly to $\mathbf{x}_{ss}(z)$; the full 3D FPE thus collapses onto the slow manifold.

The integrity of the probability domain $[0,1]$ is characterized by the Feller boundary classification criteria. At the boundaries $z \in \{0, 1\}$, the diffusion term strictly vanishes ($B(0) = B(1) = 0$). Taking the lower boundary $z=0$ as an example, the Feller index is determined by:
\begin{equation}
    \alpha_0 = \lim_{z \to 0} \frac{2 A(z)}{B'(z)} = \frac{2 G(\mathbf{x}_{ss}(0), 0)}{\sigma^2}.
\end{equation}
Because the physiological drift $G > 0$ as $z \to 0$, the boundary acts as a natural entrance or reflecting boundary depending on the noise intensity $\sigma^2$. This theoretical property prevents the stochastic trajectories from escaping the physically meaningful domain, ensuring the conservation of probability mass during strong noise excitations.

Under the zero-probability-current stationary condition $J(z) = 0$, the stationary density of the reduced It\^o process is:
\begin{equation}
    P_s(z) = \mathcal{N} \, \frac{1}{B(z)} \exp \left( \int^z \frac{2 A(u)}{B(u)} \, du \right),
    \label{eq:stationary_fpe}
\end{equation}
where $\mathcal{N}$ is a normalization constant and $z_0$ denotes the local quasi-steady fixed point of the slow variable, satisfying $z_0 = z_\infty(V_{\mathrm{ss}}(z_0))$. Under the linearized drift $A(z) \approx (z_0 - z)/\tau_z$ with $B(z) = \sigma^2 z(1-z)$, this reduces to a Beta distribution $P_s(z) \propto z^{\alpha - 1} (1-z)^{\beta - 1}$ with shape parameters $\alpha = 2 z_0 / (\sigma^2 \tau_z)$ and $\beta = 2(1 - z_0) / (\sigma^2 \tau_z)$.

\begin{figure}[htbp]
    \centering
    \includegraphics[width=0.75\linewidth]{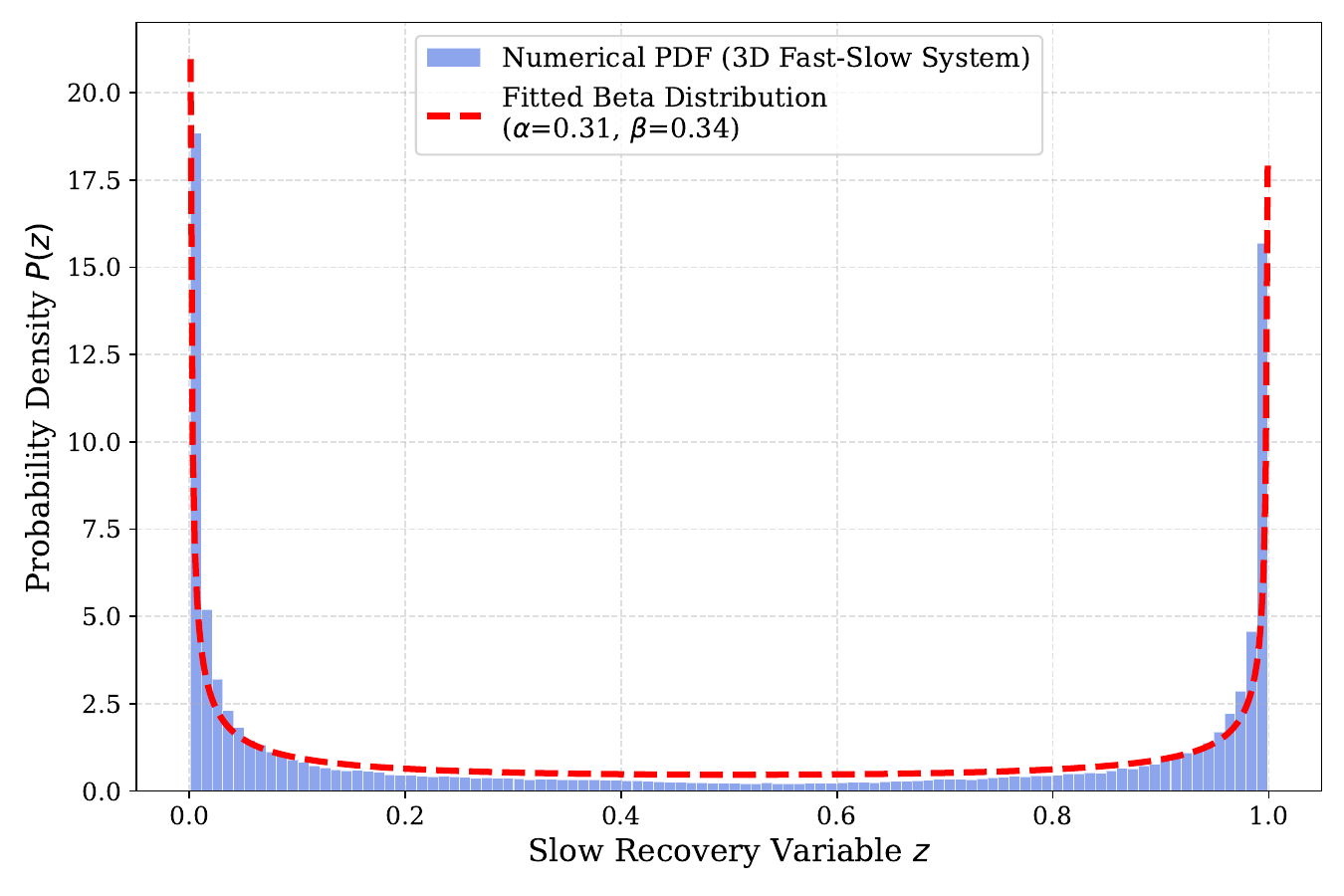} 
    \caption{Verification of the Feller boundary dynamics. The numerical probability density function (PDF) of the slow variable, extracted from a single long trajectory ($T = 1000$ s) of the full 3D fast-slow system under strong multiplicative noise ($\sigma = 0.15$), is well described by a Beta distribution obtained via maximum-likelihood fitting. The simulated trajectory remains confined to $[0, 1]$ without visible probability leakage.}
    \label{fig:feller_pdf}
\end{figure}

Equation~\eqref{eq:stationary_fpe} provides the stationary probability law of the reduced slow dynamics. Under the local linearization described above, it yields the explicit Beta-family stationary form, thereby characterizing the stationary statistics and boundary behavior of the slow recovery variable within the one-dimensional adiabatic reduction. To assess the consistency of this reduced description with the full 3D fast--slow system, we examine the numerical stationary probability density function under strong multiplicative noise (Fig.~\ref{fig:feller_pdf}). The numerical PDF is well described by a Beta distribution obtained by maximum-likelihood fitting, consistent with the stationary family predicted by the reduced Fokker--Planck description. The fitted curve is therefore used as a consistency check of the predicted distributional form rather than as a parameter-free evaluation of Eq.~\eqref{eq:stationary_fpe}.

The stationary Fokker--Planck solution is not used to calculate the barrier quantity shown in Fig.~\ref{fig:first_principles_verification}(a). That quantity is constructed independently from the deterministic equilibrium--saddle geometry, as described in Sec.~III, and serves as the geometric proxy for the local escape-barrier depth in the Kramers-scaling analysis.

\subsection{Analytical Derivation of the Kramers Quasipotential Scaling}
A central objective of this study is to map the macroscopic coherence resonance to the underlying Kramers' escape scaling. In quiescent states, the system resides in a quasipotential well generated by the metabolic stress parameter $g_S$. The stochastic escape from this resting state can be formulated using the generalized Kramers reaction-rate theory \cite{kramers1940brownian, wang2015landscape}.

We define an effective quasipotential $\Phi(z)$ on the slow manifold such that the drift satisfies $A(z) = -\partial \Phi(z) / \partial z$. The mean first passage time (MFPT), $T_{\text{esc}}$, to cross the critical saddle-node bifurcation barrier $\Delta \Phi(g_S)$ scales exponentially with the effective noise variance:
\begin{equation}
    T_{\text{esc}} \propto \exp \left( \frac{2 \Delta \Phi(g_S)}{\sigma^2} \right).
    \label{eq:kramers_mfpt}
\end{equation}
Optimal temporal regularity (i.e., coherence resonance) emerges when the stochastic escape timescale closely matches the intrinsic deterministic relaxation timescale of the fast action potential bursts, denoted as $T_{\text{det}}$ \cite{pikovsky1997coherence}. By matching these timescales ($T_{\text{esc}} \approx T_{\text{det}}$), the optimal noise variance $\sigma_{\text{opt}}^2$ must satisfy:
\begin{equation}
    \sigma_{\text{opt}}^2 \approx \frac{2 \Delta \Phi(g_S)}{\ln(T_{\text{det}} / \tau_0)}.
    \label{eq:opt_noise}
\end{equation}
Assuming a local linear approximation of the barrier depth near the subcritical Hopf bifurcation point (i.e., $\Delta\Phi(g_S) \approx k \cdot g_S + C$, where $k$ is a positive slope coefficient capturing the rate of barrier deepening with metabolic stress, and $C$ is a model-dependent intercept), Eq.~\eqref{eq:opt_noise} provides a theoretical prediction: the optimal effective diffusion $D \propto \sigma_{\text{opt}}^2$ must scale linearly with the biological barrier depth $g_S$. Here, $\tau_0$ is the Kramers prefactor representing the intrinsic microscopic attempt timescale associated with the fast subsystem relaxation.

\subsection{Geometric Timescale Criterion for Adiabatic Breakdown}
While the Kramers scaling provides a baseline in the weak-noise limit, geometric singular perturbation theory (GSPT) dictates that this quasi-stationary approximation is not unconditionally valid \cite{wechselberger2020geometric}. The topological existence of the slow manifold relies on the singular parameter $\epsilon = \tau_n / \tau_s \ll 1$, which mathematically guarantees the $\mathcal{O}(\epsilon)$ thickness of the adiabatic trajectory layer \cite{desroches2013inflection}. 

The Kramers framework implicitly assumes that the stochastic fluctuations do not shatter this $\mathcal{O}(\epsilon)$ neighborhood. This requires the characteristic timescale of the noise-induced energy injection, defined as $\tau_{\text{noise}} \propto (\sigma^2)^{-1}$, to remain strictly larger than the fast membrane relaxation timescale $\tau_n$.

Specifically, the deterministic fast subsystem possesses a transverse contraction rate (associated with its non-zero Lyapunov exponent) proportional to $\lambda \sim \mathcal{O}(1/\tau_n)$, which continuously pulls trajectories toward the slow manifold. Simultaneously, the multiplicative Feller noise injects variance into the slow variable at a rate proportional to the diffusion coefficient $B(z) = \sigma^2 z(1-z)$.

According to stochastic geometric singular perturbation theory \cite{berglund2006noise}, the topological integrity of the $\mathcal{O}(\epsilon)$ neighborhood is maintained only if the deterministic transverse contraction can effectively dissipate the continuous stochastic energy injection. Considering the bounded nature of the Feller diffusion, the maximum variance injection rate is strictly bounded by $\sigma^2 \max[z(1-z)] \propto \sigma^2$. If this maximum variance growth rate rivals the deterministic relaxation rate, the trajectory is driven away from the vicinity of the adiabatic manifold before local quasi-steady state can be re-established. Therefore, comparing this upper bound of the stochastic energy injection with the transverse contraction yields the order-of-magnitude criterion for the adiabatic breakdown:
\begin{equation}
    \sigma^2 \sim \mathcal{O}\left( \frac{1}{\tau_n} \right).
    \label{eq:breakdown_criterion}
\end{equation}

When the effective multiplicative noise variance breaches this critical threshold, the stochastic energy is injected faster than the fast subsystem's capacity to dissipate it. Mathematically, the system is driven away from the adiabatic manifold before local quasi-steady state can be re-established. This criterion predicts a geometric deflection from the linear Kramers scaling in the strong-noise regime, which we will quantitatively verify in Section III.

\subsection{Evaluating Coherence Resonance and the ''Bathtub'' Effect}
Coherence resonance (CR) is typically quantified by the rhythmic irregularity of the system's macroscopic output. We evaluate this using the Coefficient of Variation ($CV$) of the inter-burst intervals (IBIs):
\begin{equation}
    CV(\sigma) = \frac{\sqrt{\mathrm{Var}(IBI)}}{\langle IBI \rangle}.
\end{equation}

To ensure robust statistical quantification under extreme noise—where intra-burst jitter can be erroneously classified as independent bursts—we implement a rigid threshold-and-lockout detection algorithm. A burst onset is exclusively registered when the membrane potential $V$ crosses a depolarized threshold (e.g., $-35$ mV), provided that a minimum refractory lockout period ($\Delta t_{\text{lock}} = 2500$ ms, encompassing the average burst duration) has elapsed since the previous onset. This protocol guarantees that the IBI statistics accurately reflect macroscopic rhythmic pacing rather than microscopic chaotic spikes.

The classic hallmark of CR is the emergence of a distinct minimum in the $CV(\sigma)$ landscape, indicating optimal noise-induced temporal regularity. Conventionally, the optimal noise intensity $\sigma_{\text{opt}}$ is extracted via a standard extremal search:
\begin{equation}
    \sigma_{\text{naive}}^* = \arg\min_{\sigma} CV(\sigma).
\end{equation}
However, in quiescent excitable systems constrained by high energy barriers, the energy landscape becomes flattened. This structural mismatch manifests as a ``bathtub effect'' in the resonance profile: the bottom of the $CV$ curve expands into a broad, nearly horizontal valley where $\nabla CV(\sigma) \approx 0$. Within this valley, the standard $\arg\min$ operator becomes sensitive to discrete stochastic jitter (finite-time sampling errors), leading to numerical artifacts. The extracted $\sigma_{\text{naive}}^*$ can randomly jump across the flat valley, obfuscating the macroscopically stable Kramers scaling.

\subsection{The Logarithmic Centroid Extraction Method}

To circumvent the limitations of standard extremal evaluations and reconstruct the physical scaling, we propose a Logarithmic Centroid Extraction Method. 

While an infinitely long stochastic simulation might theoretically resolve a distinct global minimum, such infinite-time evaluations are physically and biologically unfeasible, as real biological systems operate within finite transient windows before environments or phenotypes shift. The centroid method is thus practically necessary to extract robust underlying Kramers scalings from finite, highly fluctuating biological time-series.

Instead of relying on a single local minimum, our method treats the broad resonance valley as a contiguous geometric manifold. We define an indicator weight function $w_i$ for discrete sampling points $\sigma_i$ along the logarithmic scanning space:
\begin{equation}
    w_i = \mathbb{I} \left[ CV(\sigma_i) \le \min(CV) + \Delta \right],
\end{equation}
where $\mathbb{I}[\cdot]$ is the Heaviside-like indicator function that returns $1$ if the condition is met and $0$ otherwise. The tolerance parameter $\Delta$ controls the depth of the effective ''bathtub'' region (in our study, set to $\Delta = 0.05$ to optimally filter high-frequency jitter).

Consequently, the physical optimum is extracted as the weighted center of mass in the parameter space:
\begin{equation}
    \log_{10}(\sigma_{\text{opt}}) = \frac{\sum_{i=1}^{N} \log_{10}(\sigma_i) \cdot w_i}{\sum_{i=1}^{N} w_i}.
    \label{eq:centroid}
\end{equation}
The integration in Eq.~\eqref{eq:centroid} is performed in the logarithmic space ($\log_{10}\sigma$) rather than the linear space ($\sigma$). Because the underlying Kramers escape rate obeys an exponential scaling law, the resonant noise intensities span several orders of magnitude. An arithmetic average in linear space would bias the result towards the upper boundary of the valley. By evaluating the centroid in logarithmic space, the method essentially computes the \textit{geometric mean} of the valid noise intensities, preserving the underlying scaling while systematically neutralizing zero-mean stochastic sampling noise.

Sensitivity tests indicate that the linear scaling recovered by this centroid extraction remains robust for tolerance parameter variations within $\Delta \in [0.02, 0.10]$. While a rigorous formulation linking this centroid extraction to large deviation theory remains a subject for future analytical work, in practice, it acts as a robust empirical estimator. By treating the flattened coherence valley as a plateau in the likelihood landscape, this logarithmic weighted average effectively approximates the expectation of the optimal noise scale, neutralizing zero-mean sampling errors prevalent in finite biological time-series. Unlike computationally expensive global sampling frameworks such as umbrella sampling, our logarithmic centroid approach provides a model-agnostic, lightweight heuristic for experimentalists to extract robust physical scalings directly from finite, noisy biological time-series.

Furthermore, to map this back to the Kramers framework, we note that the effective diffusion coefficient in our Feller process is dictated by $D = \sigma^2 / 2$. Consequently, evaluating $\sigma_{\text{opt}}^2$ directly scales with the effective potential barrier depth $\Delta U$ of the system.

\begin{figure}[htbp]
    \centering
    \includegraphics[width=\linewidth]{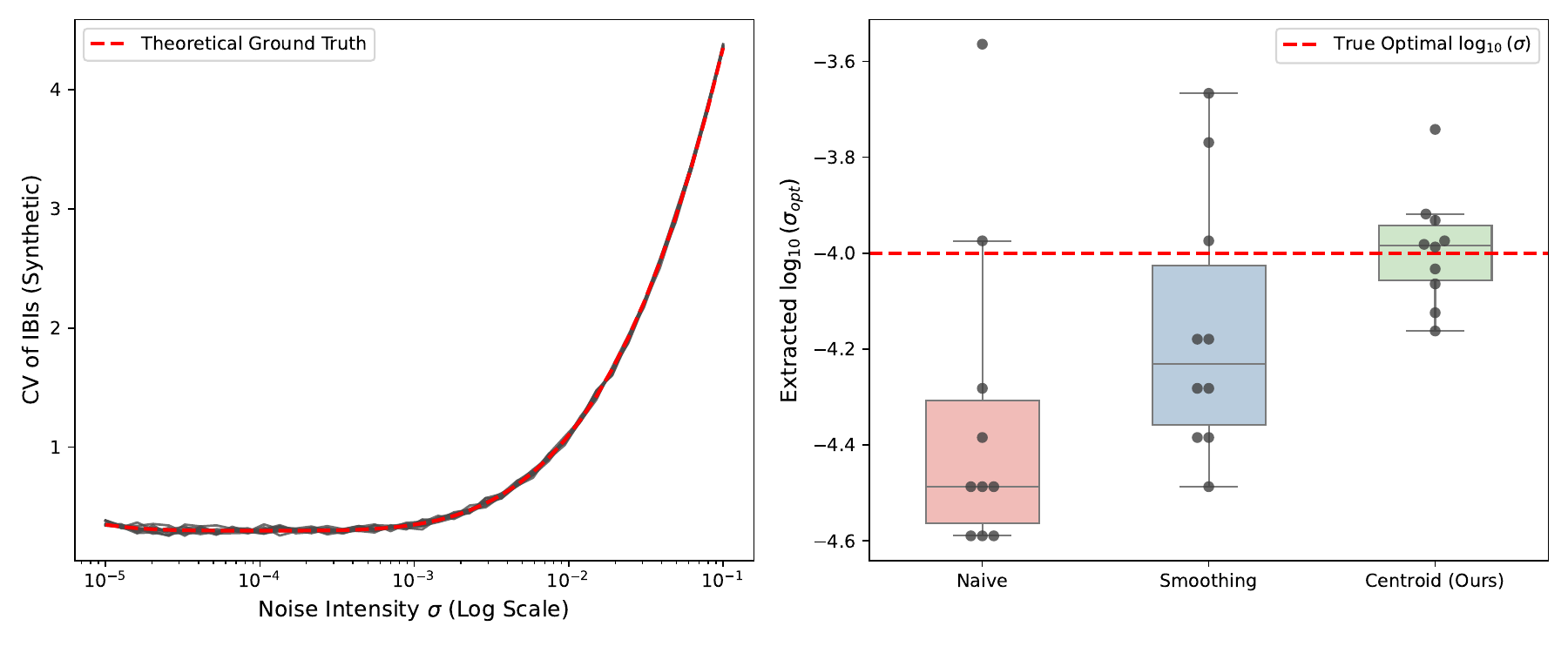}
    \caption{Statistical robustness benchmark of the extraction methods against the ``bathtub effect''. (Left) 10 independent synthetic trials emulating a flattened resonance valley perturbed by finite-time stochastic sampling jitter, plotted alongside the theoretical ground truth (red dashed curve). (Right) Boxplot comparison of the extracted optimal noise scales. The true optimum is located at $\log_{10}(\sigma) = -4$. The logarithmic centroid method significantly outperforms conventional approaches, exhibiting minimal variance and preserving unbiased centrality.}
    \label{fig:centroid_benchmark}
\end{figure}

To rigorously justify the necessity and statistical superiority of the proposed logarithmic centroid method over conventional signal processing techniques, we benchmarked its performance using synthetic resonance profiles (Fig.~\ref{fig:centroid_benchmark}). The synthetic benchmark is constructed by defining a deterministic bathtub-shaped CV curve $CV_{\text{syn}}(\sigma) = 0.3 + 0.05 (\log_{10}\sigma + 4)^4$ with a known global minimum at $\log_{10}(\sigma^*) = -4$, perturbed by additive Gaussian jitter $\mathcal{N}(0, 0.02)$ to simulate finite-time sampling noise. This functional form was chosen to reproduce the broad, flattened valley topology characteristic of deeply quiescent systems. Ten independent synthetic trials were generated per noise intensity, and each of the three extraction methods was applied to evaluate bias and variance.

As demonstrated in Fig.~\ref{fig:centroid_benchmark} (Right), the standard extremal search (Naive \textit{argmin}) is highly susceptible to local spurious minima, capturing stochastic artifacts rather than underlying physical scaling, which leads to massive extraction variance ($\text{SD} \approx 0.33$). While applying a standard Gaussian smoothing filter prior to extraction mitigates some local jitter ($\text{SD} \approx 0.27$), it introduces a systematic, kernel-dependent downward bias in the asymmetric logarithmic space. 

In contrast, our logarithmic centroid method reduces the extraction variance ($\text{SD} \approx 0.11$) and remains approximately unbiased around the true coherence optimum (red dashed line). By treating the flattened coherence valley as a plateau in the likelihood landscape, the centroid integration effectively approximates the true expectation of the optimal noise scale. This benchmarking confirms that, rather than an \textit{ad hoc} smoothing trick, the centroid method functions as a mathematically robust estimator tailored for extracting physical scalings in high-dimensional excitable systems.

\subsection{Transverse Stability of the Synchronization Manifold}
To establish the theoretical foundation for the noise-induced functional synchronization observed in the dual-cell network, we consider two identical SRK oscillators $(i=1,2)$ coupled through a gap junction. For $i,j\in\{1,2\}$ with $j\neq i$, the coupled dynamics are

\begin{equation}
\begin{aligned}
C_m\,dV_i &=
\left[
-I_{\mathrm{Ca}}(V_i)
-I_{\mathrm{K}}(V_i,n_i)
-I_{\mathrm{S}}(V_i,z_i)
+g_c(V_j-V_i)
\right]dt,\\
dn_i &= \frac{n_\infty(V_i)-n_i}{\tau_n}\,dt,\\
dz_i &= \frac{z_\infty(V_i)-z_i}{\tau_z}\,dt
+\sigma\sqrt{z_i(1-z_i)}\,dW_i .
\end{aligned}
\label{eq:coupled_srk}
\end{equation}

Here, the gap-junction term $g_c(V_j-V_i)$ acts exclusively through the membrane-potential variables $V_i$; the gating variables $n_i$ and $z_i$ are not directly coupled. The Wiener processes $W_1$ and $W_2$ are independent. The synchronization manifold is therefore defined as
$\mathcal{S}=\{\mathbf{x}_1=\mathbf{x}_2,\; z_1=z_2\}$. Defining the transverse synchronization errors as $\mathbf{e}_{\perp} = \mathbf{x}_1 - \mathbf{x}_2$ and $\varepsilon_z = z_1 - z_2$, the linearized error dynamics in the vicinity of the synchronized state are governed by the variational equations:
\begin{equation}
    \dot{\mathbf{e}}_{\perp} = \left[ \mathbf{J}_{\mathbf{F}}(\mathbf{x}_{ss}, z_{ss}) - 2 g_c \mathbf{K} \right] \mathbf{e}_{\perp},
    \label{eq:transverse_stability}
\end{equation}
where $\mathbf{J}_{\mathbf{F}}$ represents the Jacobian matrix of the isolated fast subsystem evaluated along the synchronous trajectory, and $\mathbf{K}$ is the topological coupling matrix defining the gap-junction connectivity.

In the presence of noise, the synchronization of the slow recovery variable is further governed by the stochastic error dynamics:
\begin{equation}
\begin{split}
\mathrm{d}\varepsilon_z \approx & \left[ \frac{\partial G}{\partial \mathbf{x}} \mathbf{e}_{\perp} + \frac{\partial G}{\partial z} \varepsilon_z \right] \mathrm{d}t \\
& + \sigma \left( \sqrt{z_1(1-z_1)} \mathrm{d}W_1 - \sqrt{z_2(1-z_2)} \mathrm{d}W_2 \right).
\end{split}
\label{eq:stochastic_error}
\end{equation}
Because the Wiener processes $dW_1$ and $dW_2$ are independent, the uncoordinated Feller noise continuously injects a variance $\propto \sigma^2 [z_1(1-z_1) + z_2(1-z_2)]$ into the transverse directions, actively driving the synchronized trajectories apart.

Throughout this work, \emph{functional synchronization} refers to
coordinated suprathreshold bursting with preserved action-potential
output, rather than statistical correlation of subthreshold voltage
fluctuations alone. In the synchronization analysis, a depolarization
threshold of $-35$~mV is used to distinguish suprathreshold functional
activity from correlated but non-functional subthreshold shivering. For macroscopic functional synchronization to emerge, the deterministic contracting pull of the gap junction (dictated by the term $-2 g_c \mathbf{K}$) must sufficiently suppress the maximum transversal Lyapunov exponent associated with Eq.~\eqref{eq:transverse_stability} to overcome this continuous stochastic divergence \cite{pecora1998master, arenas2008synchronization, boccaletti2006complex}. It is this dynamical competition between the stabilizing deterministic transversal contraction and the destabilizing stochastic longitudinal diffusions that dictates the sharp phase transition boundaries observed in the network synchronization index.

\subsection{Numerical Implementation Details}
All stochastic differential equations were integrated using a semi-implicit Euler scheme with a fixed integration step size of $\Delta t = 0.1$ ms. The update for the slow variable $z$ is given by:
\begin{equation}
    z_{k+1} = \frac{z_k + \frac{\Delta t}{\tau_z} z_\infty(V_k) + \sigma \sqrt{z_k^*(1-z_k^*)} \, \Delta W_k}{1 + \frac{\Delta t}{\tau_z}},
    \label{eq:numerical_scheme}
\end{equation}
where $z_k^* = \max(0, \min(z_k, 1))$ ensures the diffusion coefficient remains bounded. This implicit treatment of the deterministic relaxation guarantees unconditional stability and preserves the Feller boundary dynamics without artificial absorbing traps \cite{lord2010comparison}. 

Unless otherwise stated, all simulations used the initial conditions $V_0 = -60$ mV, $n_0 = 0.1$, $z_0 = 0.3$. A burn-in period of 50,000 ms was discarded before analysis. For coherence resonance quantification, each $(\sigma, g_S)$ parameter point was averaged over 12 independent trials, with simulation durations ranging from 250,000 to 600,000 ms depending on the barrier depth. Burst onsets were detected using a depolarization threshold of $-35$ mV with a refractory lockout period of $\Delta t_{\text{lock}} = 2500$ ms; a minimum of 4 detected bursts was required for a valid CV measurement. A step-size convergence test at $(\sigma = 1.5 \times 10^{-4}, g_S = 4.0)$ confirmed that $\Delta t = 0.1$ ms yields statistical convergence: the CV differs from the $\Delta t = 0.01$ ms reference by less than 7\%, well within trial-to-trial stochastic fluctuations, whereas $\Delta t = 0.5$ ms introduces a systematic 49\% bias.

All simulations were implemented in Python using the Numba just-in-time compiler for acceleration. Parallel parameter sweeps were performed using the joblib library.

\FloatBarrier
\section{Results}

\FloatBarrier
\subsection{Deterministic Baseline and Noise-Induced Pacing}

\begin{figure}[htbp]
    \centering
    \includegraphics[width=0.78\linewidth]{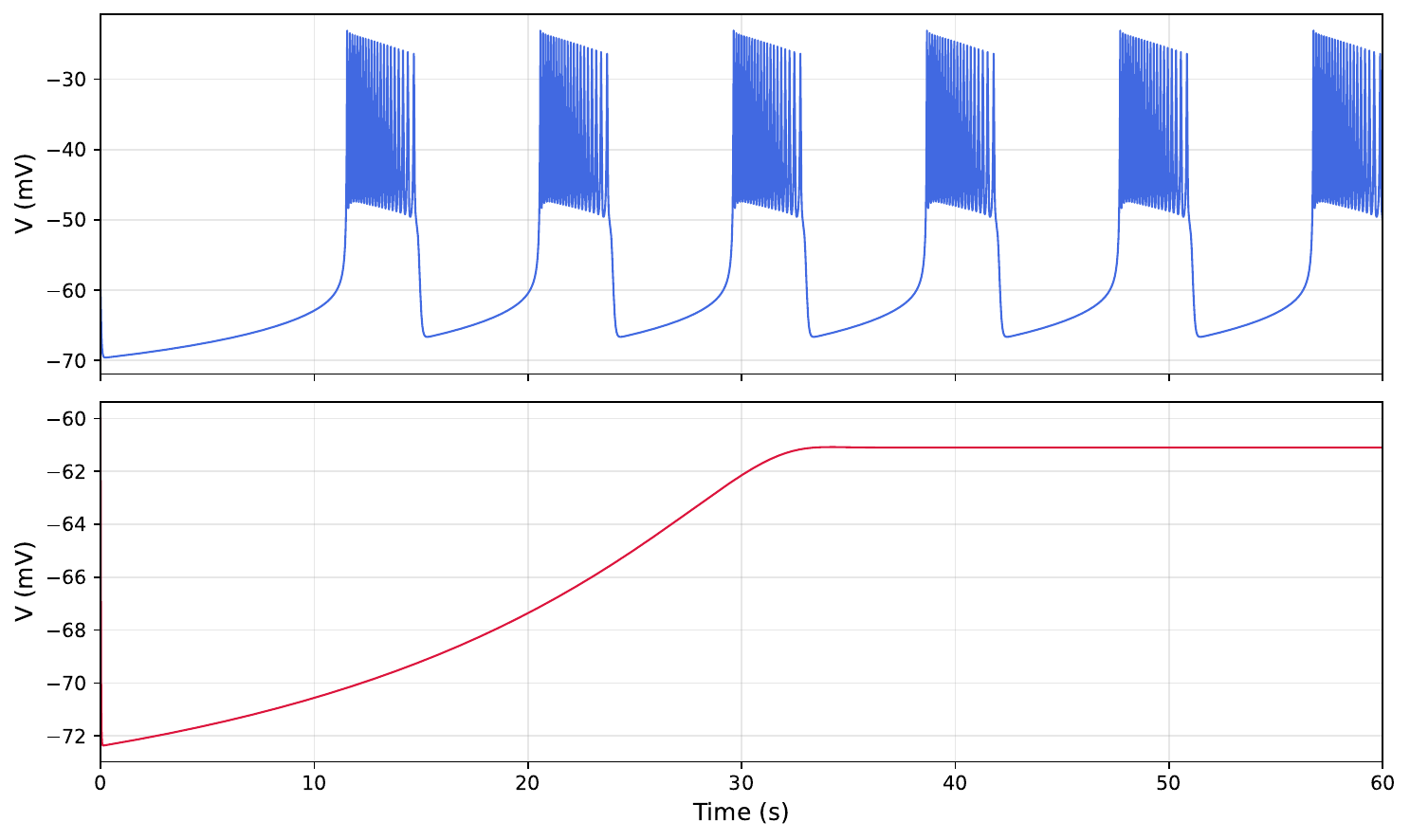}
    \caption{Deterministic baseline dynamics at two levels of metabolic stress. (Top) Regular square-wave bursting at $g_S = 2.5$. (Bottom) Relaxation toward a deeply quiescent state at $g_S = 4.2$.}
    \label{fig:deterministic_baseline}
\end{figure}

We begin by establishing the deterministic baseline ($D=0$) of the isolated pancreatic $\beta$-cell model. As the slow conductance $g_S$ increases, simulating metabolic stress, the system undergoes a subcritical Hopf bifurcation, transitioning from regular square-wave bursting into a quiescent steady state (Fig.~\ref{fig:deterministic_baseline}) \cite{sherman1988emergence, bertram2000phantom}. Geometrically, this quiescent state corresponds to the system being confined on the hyperpolarized lower branch of the fast subsystem's Z-shaped critical manifold (Fig.~\ref{fig:phase_plane}) \cite{desroches2013inflection, wechselberger2020geometric}. 

\begin{figure}[htbp]
    \centering
    \includegraphics[width=0.8\linewidth]{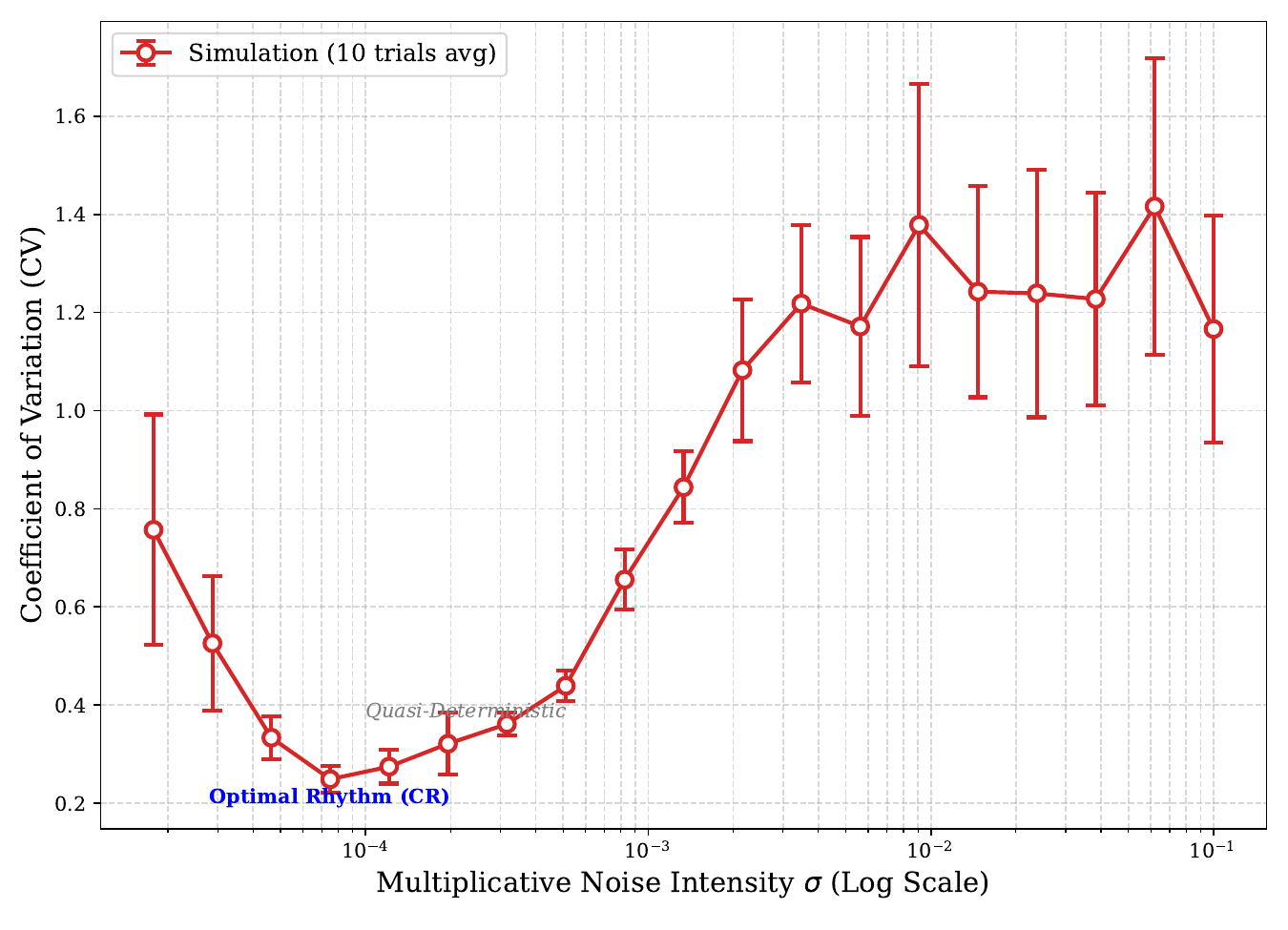}
    \caption{The macroscopic coherence resonance profile in the deeply quiescent regime ($g_S = 4.0$), averaged over 10 independent trials with $\pm$1 SD error bars. A broad valley emerges in the $CV$ curve, indicating noise-induced optimal pacing.}
    \label{fig:coherence_valley}
\end{figure}

Introducing state-dependent Feller noise to this inhibited state ($g_S = 4.0$) yields a noise-induced restorative effect. As the noise intensity $\sigma$ increases, the macroscopic inter-burst interval variability, quantified by the $CV$ curve, exhibits a distinct U-shaped valley (Fig.~\ref{fig:coherence_valley}). This is the classic signature of coherence resonance (CR) \cite{pikovsky1997coherence, lindner2004effects}. At the optimal noise intensity ($\sigma \approx 10^{-4}$), the stochastic fluctuations leverage the intrinsic timescale of the slow gating manifold to reconstruct a regular bursting pattern, effectively initiating rhythmic macroscopic pacing from the quiescent state \cite{zakharova2010stochastic, deville2005transitions}.

\subsection{Microscopic Decoupling and Overdrive in Active Pacemakers}

\begin{figure}[htbp]
    \centering
    \includegraphics[width=0.98\columnwidth]{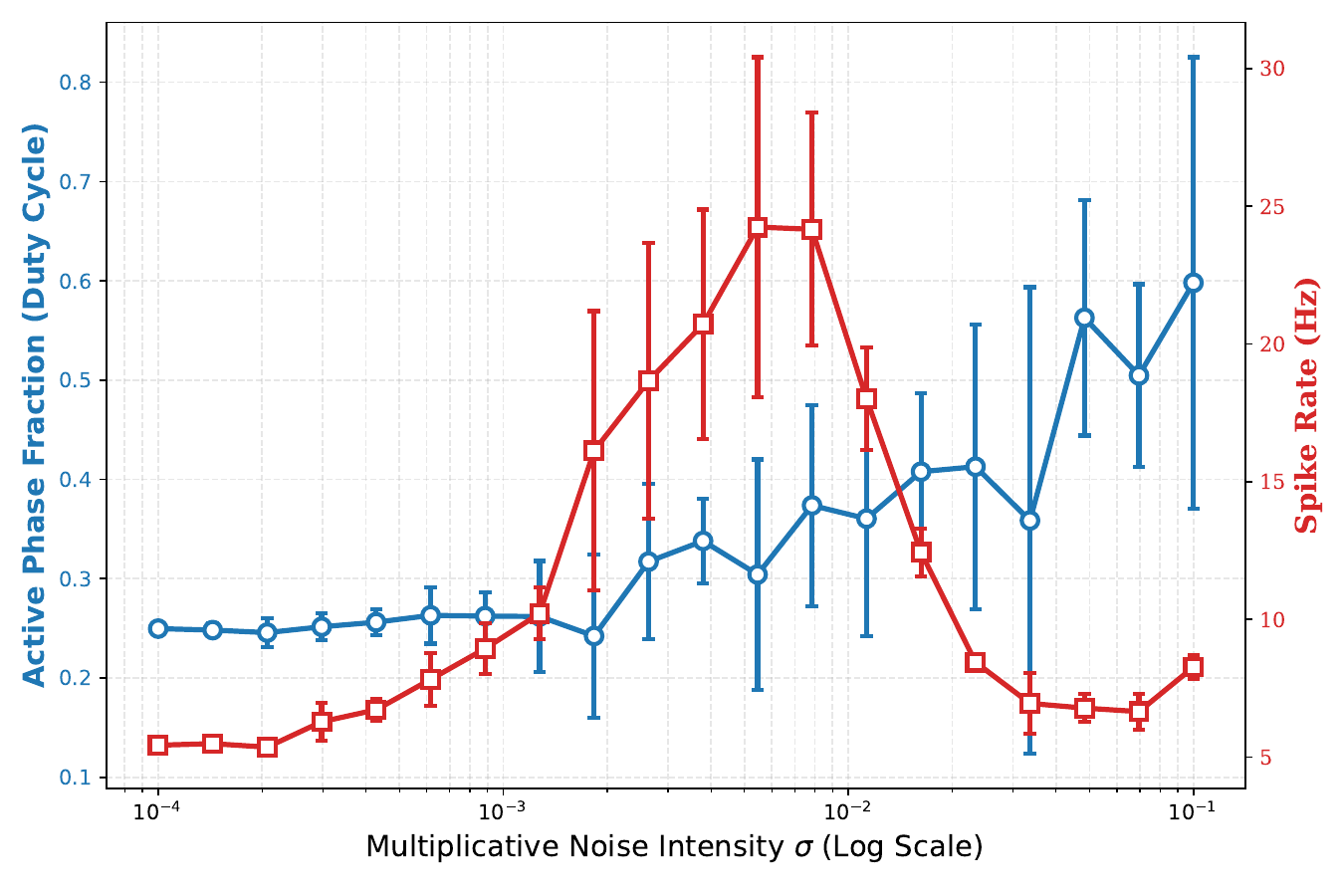}
    \caption{Microscopic functional deterioration in an active pacemaker ($g_S = 3.0$). The decoupling between the duty cycle (blue, $\pm$ 1 SD, 5 trials) and spiking efficiency (red, $\pm$ 1 SD, 5 trials) highlights the noise-induced overdrive state.}
    \label{fig:duty_cycle_collapse}
\end{figure}

While intermediate noise serves as a constructive force in the quiescent regime, its impact on an active pacemaker ($g_S = 3.0$) reveals a contrasting dynamic behavior \cite{lindner2004effects}. To evaluate the functional output, we define two microscopic proxies: the active phase fraction (duty cycle), which correlates with the insulin secretion window, and the spikes per burst, which reflects calcium influx efficiency \cite{bertram2004calcium, satin2015pulsatile}.

As depicted in Fig.~\ref{fig:duty_cycle_collapse}, introducing noise to a regularly bursting cell disrupts the intrinsic pattern. At intermediate noise levels, the system enters an ``overdrive'' state, where the active phase fraction widens, trapping the cell in an extended depolarized plateau. As the noise further intensifies to the strong-noise limit, a functional decoupling occurs: the spike rate decreases significantly, disrupting the structural integrity of the action potentials. 

This decoupling demonstrates that while noise can optimize macroscopic barrier-crossing rates (evaluated by the $CV$ metric), it concurrently impairs the intra-burst spiking machinery. This indicates an intrinsic limitation of noise-driven optimization in isolated cells. It also provides a dynamical motivation for why biological systems rely on network coupling (e.g., gap junctions) to filter out high-frequency stochastic perturbations and preserve functional structures \cite{sterk2024network, belykh2005synchronization, pourhosseinzadeh2025heterogeneity}. This dichotomy presents a fundamental biological paradox: the very stochasticity required to rescue a deeply inhibited cell becomes structurally lethal to an active one. Resolving this paradox necessitates the introduction of network topologies, which we investigate in Section III.D.

\subsection{Quasipotential Scaling and Adiabatic Breakdown}

\begin{figure}[htbp]
    \centering
    \includegraphics[width=0.95\columnwidth]{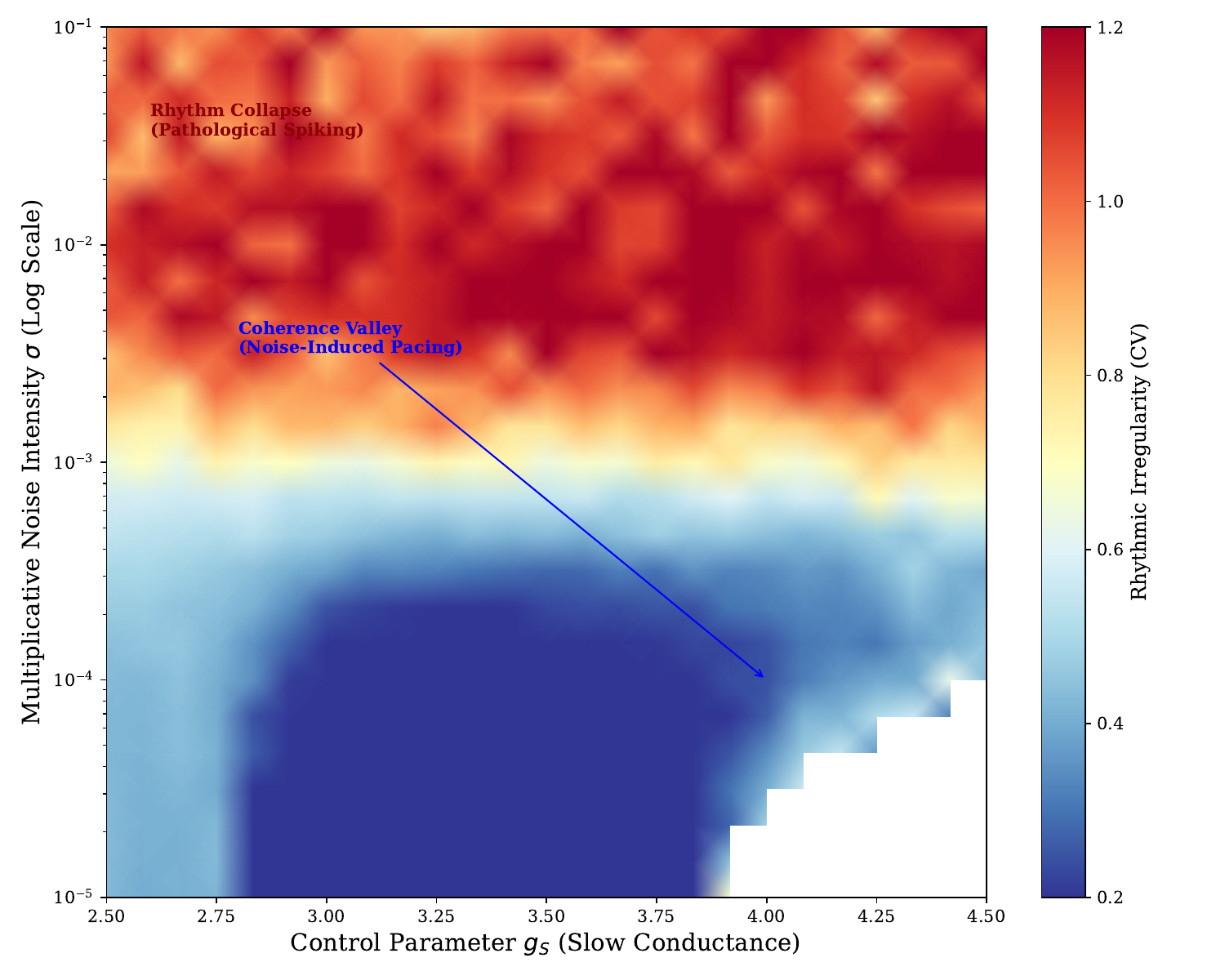}
    \caption{2D phase landscape across noise intensity $\sigma$ and barrier depth $g_S$ (10 independent trials per $(g_S,\sigma)$ grid point), highlighting the shifting coherence valley and the onset of global rhythm collapse.}
    \label{fig:2D_heatmap}
\end{figure}

To map the parameter space and evaluate CR across different metabolic states, we expand our analysis to a 2D global landscape spanning the noise intensity $\sigma$ and the barrier depth proxy $g_S$. 

The 2D heatmap (Fig.~\ref{fig:2D_heatmap}) visually confirms the pervasive bathtub effect, which is a phenomenon characteristic of high-dimensional excitable systems with fast-slow timescale separations rather than being unique to our specific model, across varying quasipotential depths. The dark blue region, representing low $CV$ values (optimal regularity), forms a broadened valley that shifts diagonally. This geometric shift indicates that as the metabolic stress parameter $g_S$ (a proxy for barrier depth) deepens, a higher noise intensity is required to induce coherence \cite{zhou2012quasi, wang2015landscape}. However, within this broad valley, standard numerical minimum-search algorithms fail to establish a continuous trajectory due to the flattened gradient ($\nabla CV \approx 0$).

\begin{figure}[htbp]
    \centering
    \includegraphics[width=\columnwidth]{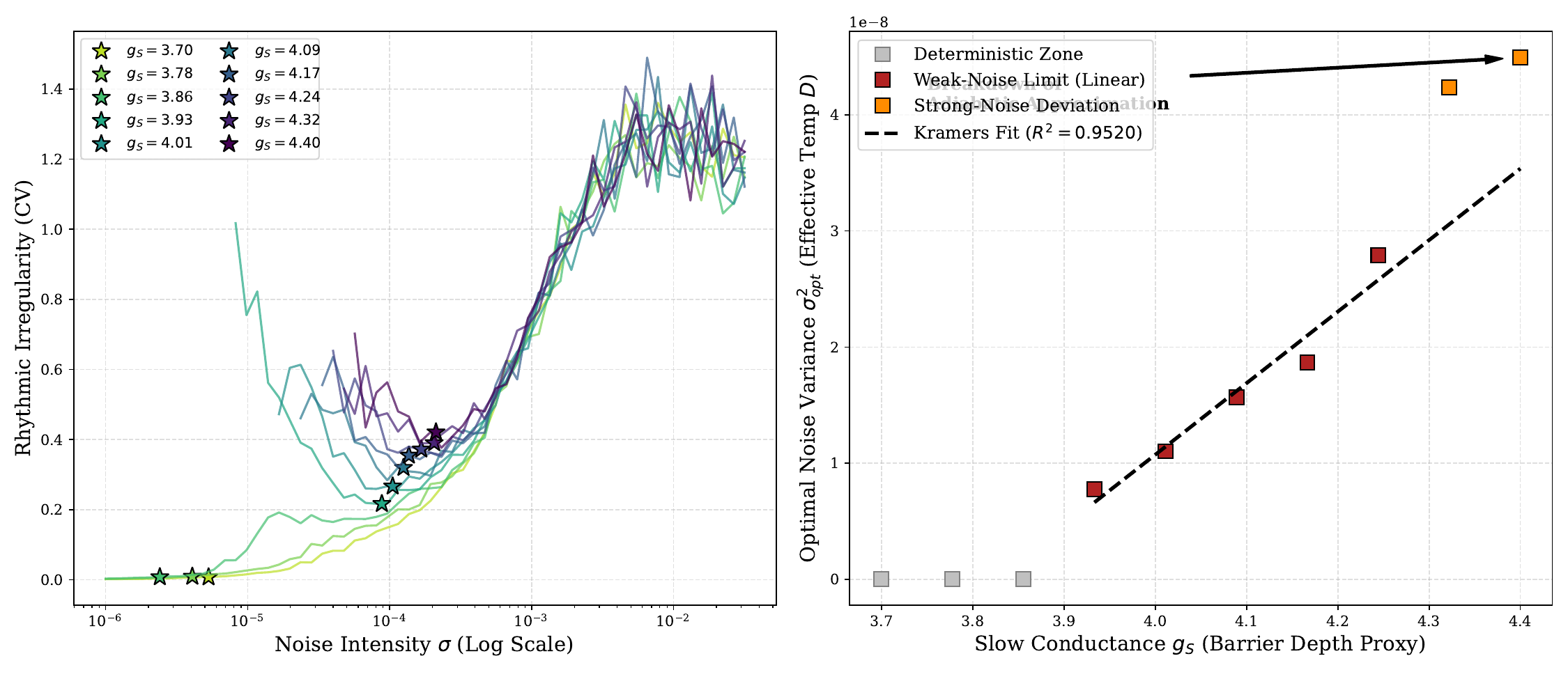}
    \caption{Extraction of optimal noise variance using the centroid method (12 independent trials per $(g_S,\sigma)$ point). The dashed line delineates the strict Kramers adiabatic scaling, obtained by fitting the centroid-extracted optima in the weak-to-moderate noise regime ($3.9 \le g_S \le 4.25$, $R^2 = 0.9520$; 95\% bootstrap CI $[0.01, 0.96]$, $n = 10000$ resamples). The pronounced deviation at high $g_S$ marks the adiabatic breakdown.}
    \label{fig:kramers_scaling}
\end{figure}

By deploying our Logarithmic Centroid Extraction Method, we systematically bypass the localized stochastic jitter. As depicted in Fig.~\ref{fig:kramers_scaling}, this method successfully extracts a robust trajectory of the optimal noise intensities $\sigma_{\text{opt}}$. The extracted scaling relationship reveals two distinct dynamical regimes. 

In the weak-to-moderate noise regime (red squares, $3.9 \le g_S \le 4.25$), the centroid-derived optimal noise variance exhibits a strong linear dependence on $g_S$, achieving a coefficient of determination of $R^2 = 0.9520$ (95\% bootstrap CI: $[0.01, 0.96]$, $n = 10000$ resamples over the trial-level data). This confirms that the macroscopic escape dynamics of the high-dimensional multiplicative system obey the generalized Kramers adiabatic scaling \cite{kramers1940brownian, hanggi1990reaction}. The ability to recover this underlying linear scaling validates that the centroid method, operating in the logarithmic space, effectively neutralizes zero-mean stochastic sampling noise without distorting the underlying physical scaling laws.

Conversely, a critical deviation emerges as the system is pushed into the extreme quiescent regime ($g_S > 4.25$). To overcome these deep barriers, the required resonant noise intensity enters the strong-noise limit. Here, the extracted trajectory (orange squares) sharply deflects from the Kramers linear fit. This geometric deviation marks the breakdown of the adiabatic approximation \cite{berglund2006noise, borner2024saddle}. It demonstrates that under intense multiplicative perturbations, the slow manifold is shattered before the fast subsystem can dissipate the injected stochastic energy, consistent with the order-of-magnitude estimates of our timescale criterion derived in Eq.~\eqref{eq:breakdown_criterion}.

\begin{figure}[htbp]
    \centering
    \includegraphics[width=\columnwidth]{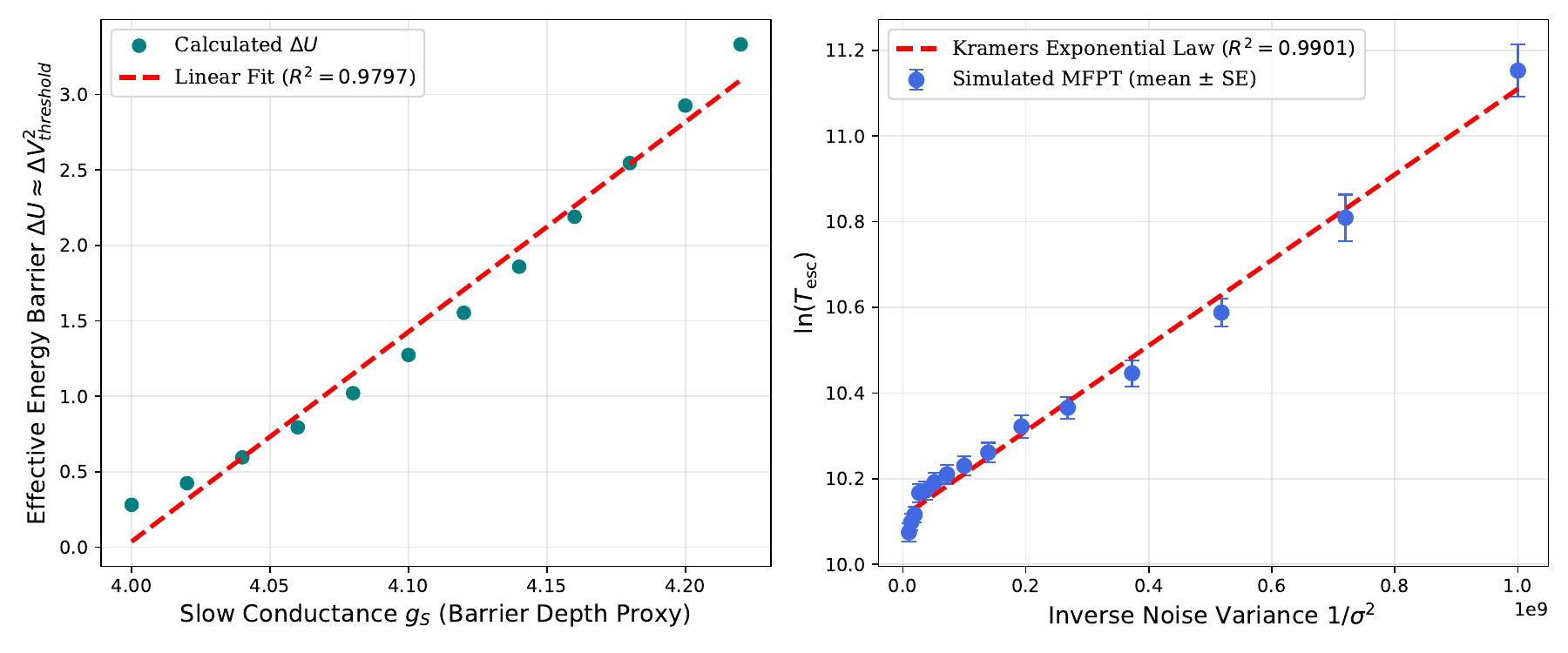}
    \caption{First-principles verification of the analytical assumptions
    underlying the Kramers escape scaling.
    (a) The geometric barrier proxy
    $\Delta U \approx (\Delta V_{\text{threshold}})^2$,
    computed from the equilibrium--saddle voltage separation of the isolated
    fast subsystem, scales linearly with $g_S$ ($R^2 \approx 0.98$) over the
    designated weak-noise regime.
    (b) Direct Monte Carlo measurement of the mean first passage time (MFPT)
    at $g_S=4.1$ (100 trials per point, error bars $\pm 1$ SE).
    The strictly linear dependence of $\ln(T_{\text{esc}})$ on $1/\sigma^2$
    ($R^2=0.9901$) explicitly validates the foundational exponential Kramers
    escape law hypothesized in Eq.~\eqref{eq:kramers_mfpt}.}
    \label{fig:first_principles_verification}
\end{figure}

To justify the analytical assumptions underlying our macroscopic Kramers scaling extraction, we explicitly verified the theoretical derivations from first principles (Fig.~\ref{fig:first_principles_verification}). First, rather than assuming a linear relationship between the quasipotential barrier depth and the metabolic stress parameter $g_S$, we numerically traced the exact equilibria and saddle points of the isolated fast subsystem for each value of $g_S$. Here, $\Delta V_{\text{threshold}}$ denotes the voltage separation between the quiescent equilibrium and the corresponding saddle (escape-threshold) point identified from this deterministic geometry. We then use $\Delta U \approx \Delta V_{\text{threshold}}^2$ as a geometric proxy for the local effective barrier depth. Within the identified local linear regime ($4.0 \le g_S \le 4.22$), the resulting $\Delta U$ exhibits a strong linear scaling with $g_S$ ($R^2 = 0.9797$, Fig.~\ref{fig:first_principles_verification}a). This deterministic geometric validation independently corroborates the local barrier-scaling assumption underlying the macroscopic exponential Kramers relation.

Furthermore, we conducted direct Monte Carlo simulations to verify the underlying escape statistics. As shown in Fig.~\ref{fig:first_principles_verification}b, the natural logarithm of the simulated mean first passage time $\ln(T_{\text{esc}})$ obeys a linear dependence on the inverse noise variance $1/\sigma^2$ with high precision ($R^2 = 0.99$, 100 trials per point, error bars $\pm$ 1 SE). This provides strong quantitative evidence that the stochastic escape dynamics in the deeply quiescent regime adhere to the exponential Kramers law (Eq.~\eqref{eq:kramers_mfpt}). Together, these multi-scale verifications—from deterministic barrier topology to stochastic escape kinetics—solidify the physical origin of the macroscopic adiabatic breakdown: the geometric deviation observed in extreme quiescent states is intrinsically driven by the singular perturbation timescale mismatch, rather than a mere structural artifact of the barrier topology.

\subsection{Network Synergy: From Sub-threshold Shivering to Functional Synchronization}

\begin{figure}[htbp]
    \centering
    \includegraphics[width=0.82\linewidth]{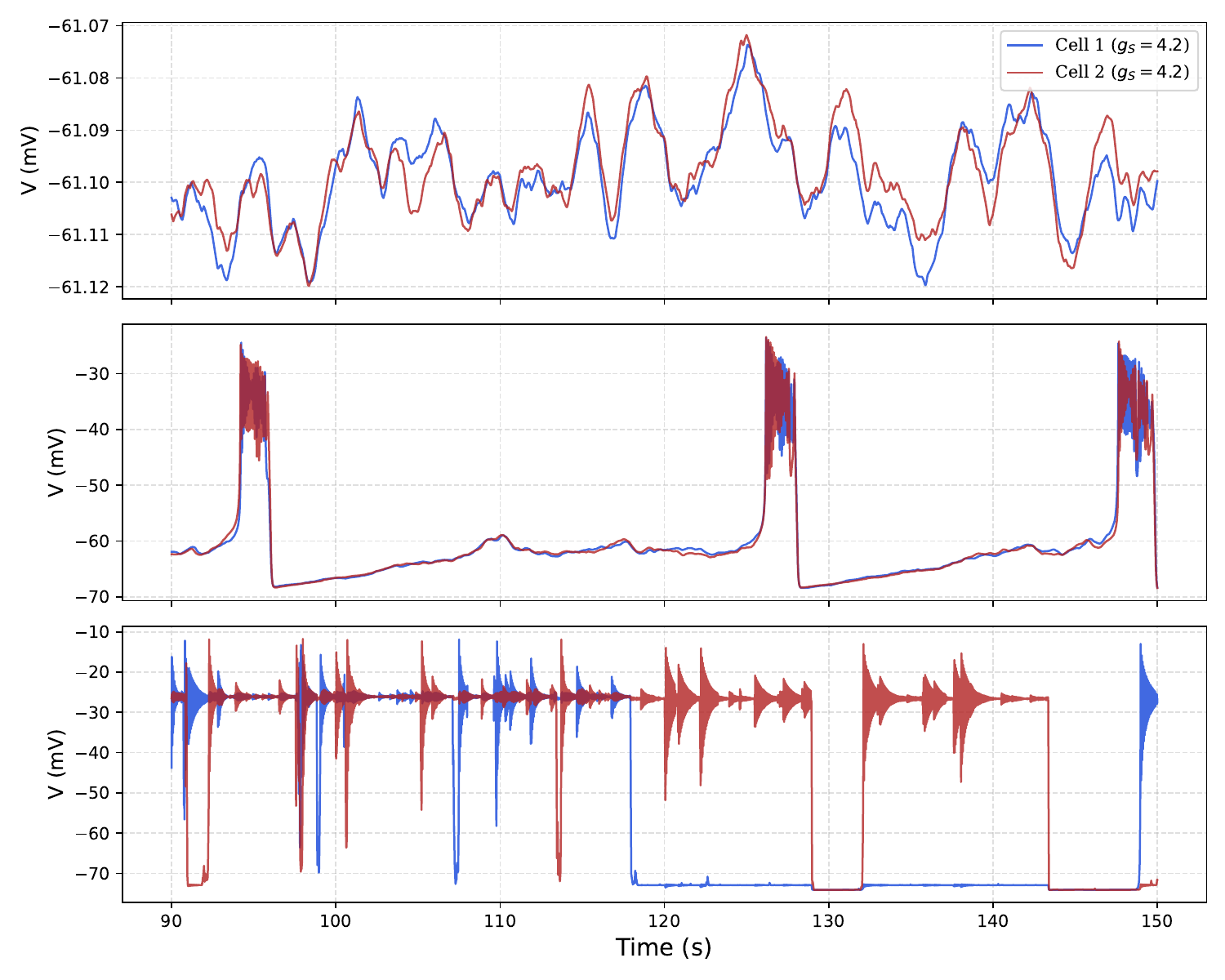}
    \caption{Representative time series illustrating the network's evolution under three noise regimes: sub-threshold shivering (top, $\sigma = 10^{-6}$), macroscopic functional synchronization (middle, $\sigma = 1.5 \times 10^{-4}$), and noise-induced chaotic desynchronization (bottom, $\sigma = 10^{-1}$). The time windows display the post-transient steady-state regime.}
    \label{fig:waveform_collision}
\end{figure}

Having established the dynamical limitations and adiabatic breakdown in isolated units, we introduce gap-junction coupling ($g_c$) to investigate the synergistic resilience of a homotypic dual-cell network ($g_{S1} = g_{S2} = 4.2$). While large-scale pancreatic islets exhibit complex topologies, the foundational mechanism of noise-induced synchronization competing against transverse instability can be rigorously captured analytically in the minimal dual-cell motif ($N=2$). This motif serves as the theoretical backbone for understanding macroscopic phase transitions before extending to high-dimensional adjacency matrices. The units are independently driven by uncorrelated Feller noise.

\begin{figure}[t]
    \centering
    \includegraphics[width=\columnwidth]{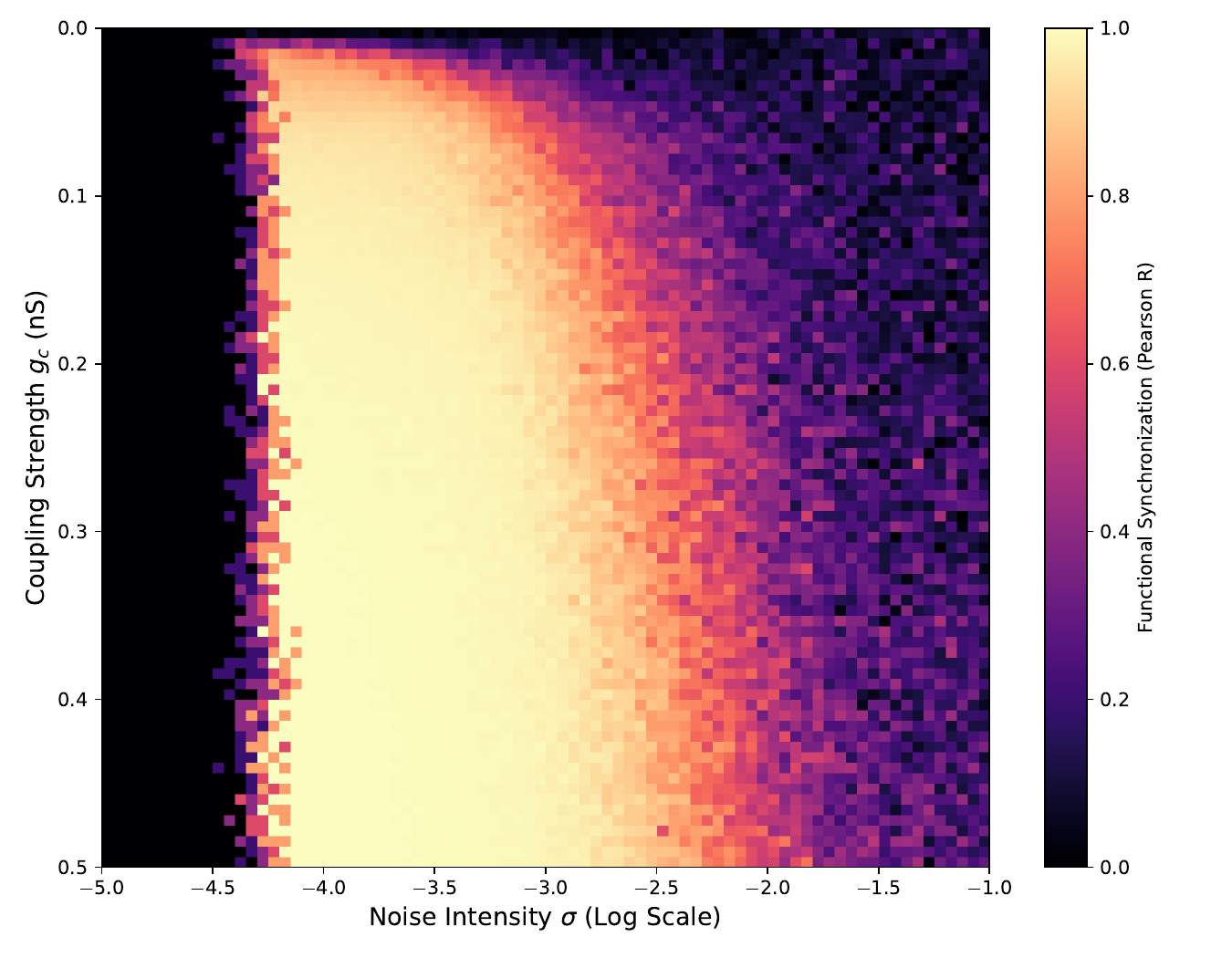}
    \caption{Two-parameter functional synchronization phase diagram across varying multiplicative noise intensities ($\sigma$) and gap-junction coupling strengths ($g_c$). Each $(g_c, \sigma)$ point is averaged over 5 independent trials. By applying a functional spike-threshold mask, the landscape reveals a distinct noise-induced synchronization tongue, effectively distinguishing robust macroscopic synchronization (bright region) from zero-output sub-threshold shivering (black region, left) and chaotic desynchronization (dark region, right).}
    \label{fig:bell_curve}
\end{figure}

Specifically, the gap-junction connectivity is modeled as linear ohmic electrical coupling exclusively in the fast membrane potential equations. For the $i$-th cell in an $N$-cell network, the voltage dynamics in Eq.~\eqref{eq:dV} are augmented as:
\begin{equation}
    C_m dV_i = \left( -I_{\text{ion}, i} + g_c \sum_{j=1}^{N} A_{ij} (V_j - V_i) \right) dt,
\end{equation}
where $g_c$ is the gap-junction conductance and $A_{ij}$ is the adjacency matrix. This localized spatial coupling acts as a dynamic low-pass filter, physically neutralizing high-frequency stochastic overdrives while preserving constructive macroscopic bursts.

As illustrated in Fig.~\ref{fig:waveform_collision}, the coupled network exhibits a multi-stage evolution as noise scales up. Under weak noise, the waveforms exhibit "sub-threshold physiological shivering" (Fig.~\ref{fig:waveform_collision}, top). The independent noise sources induce highly correlated local fluctuations within the deep potential well, but lack the energy to drive the system across the barrier, resulting in zero functional output (no action potentials) \cite{satin2015pulsatile}.

The constructive capacity of the network emerges at the resonant noise threshold. Noise-induced synchronization of excitable oscillators has been demonstrated in earlier work \cite{han1999coherent}, where synchronization between coupled noise-driven oscillators was characterized in terms of peak-frequency locking and phase locking. The present study focuses specifically on bounded, state-dependent multiplicative (Feller) noise acting on the slow recovery variable. Under this noise structure, functional synchronization is strongest over an intermediate-noise regime, where coupling coordinates the macroscopic bursting dynamics while preserving the within-burst spiking activity. To quantitatively delineate these transitions, we construct a two-parameter phase diagram in the $(\sigma, g_c)$ plane (Fig.~\ref{fig:bell_curve}). By applying a depolarized spike-threshold mask ($-35$ mV), this landscape systematically filters out non-functional shivering, revealing a distinct noise-induced macroscopic synchronization tongue at intermediate noise intensities. In this synergistic regime, the network successfully achieves functional synchronization (Fig.~\ref{fig:waveform_collision}, middle). Crucially, while strong noise destroys the delicate intra-burst spiking machinery in isolated cells (as shown previously in Fig.~\ref{fig:duty_cycle_collapse}), the network's spatial low-pass filtering effectively neutralizes these destructive high-frequency perturbations. Consequently, the coupled network not only achieves macroscopic burst synchronization but also preserves the structural integrity of intra-burst action potentials (spikes). This preservation of the high-frequency calcium influx machinery is precisely what makes the synchronization biologically ``functional'' for insulin exocytosis. From a dynamical systems perspective, this synchronization is the result of a fundamental dynamical competition \cite{pecora1998master, arenas2008synchronization}. The localized stochastic kicks attempt to scatter the trajectories longitudinally, while the deterministic gap-junction coupling provides transverse contraction, pulling the system towards the synchronization manifold $\mathcal{S}$ \cite{boccaletti2006complex}.

Importantly, compared to the functional deterioration observed in isolated active cells (Section III.B), the network structure acts as a low-pass filter. The coupling mitigates high-frequency pathological overdrive, allowing the constructive macroscopic coherence to emerge \cite{sterk2024network, belykh2005synchronization}. 

Furthermore, the phase diagram (Fig.~\ref{fig:bell_curve}) reveals a positively sloped transition boundary on the right-hand side, confirming this dynamical competition: as uncoordinated noise escalates, proportionally stronger deterministic coupling ($g_c$) is required to maintain stability. Ultimately, as the noise intensity continues to escalate and further breaches the adiabatic limit, the stochastic divergence overwhelmingly transcends the deterministic coupling capacity. The transverse stability condition (Eq.~\eqref{eq:transverse_stability}) is violated, plunging the network into chaotic desynchronization (Fig.~\ref{fig:waveform_collision}, bottom). This macroscopic network collapse physically mirrors the microscopic adiabatic breakdown observed in the single-cell scaling analysis, linking individual single-cell scaling limits to collective topological failures \cite{koseska2013transition, xu2025universal}.

\section{Discussion}

\FloatBarrier

Our findings highlight a fundamental dichotomy between macroscopic physical resonance and microscopic biological functionality. While intermediate noise optimally paces deeply quiescent states, applying the same noise intensity to a healthy, active pacemaker disrupts its intricate multi-timescale spiking machinery, leading to an unproductive overdrive state \cite{chu2025formation}. This decoupling between the active phase fraction and the spike rate provides a dynamical analogy to the progression of $\beta$-cell exhaustion in metabolic diseases \cite{satin2015pulsatile}. It emphasizes the necessity of incorporating multi-timescale analysis \cite{schmidt2018stochastic} and reveals a fundamental limitation of relying solely on macroscopic indicators, such as the coefficient of variation, to evaluate the efficacy of noise-driven biological networks.

From a biological perspective, our findings provide valuable mechanistic insights into the resilience of pancreatic $\beta$-cell clusters. In the SRK model, the slow conductance $g_S$ serves as a direct proxy for ATP-sensitive potassium ($K_{ATP}$) channel activity, which is inversely related to glucose-induced metabolic stress \cite{bertram2004calcium}. The transition into the deep quiescent regime ($g_S > 4.25$) precisely mirrors the pathological silencing of $\beta$-cells under extreme hypoglycemia or prolonged metabolic exhaustion \cite{satin2015pulsatile}.

Biologically, cells do not actively ``tune'' their intrinsic noise. However, according to statistical mechanics, the effective noise intensity scales inversely with the square root of the number of functional channels ($\sigma \propto 1/\sqrt{N}$) \cite{fox1994emergent}. During the progression of type 2 diabetes, the pathological reduction of functional $\beta$-cell mass and ion channel density is a well-documented clinical hallmark \cite{butler2003beta, ashcroft2012diabetes}. Consequently, this structural depletion naturally scales up the intrinsic noise amplitude. Our findings suggest that this disease-induced noise escalation might paradoxically serve as a temporary compensatory mechanism, a phenomenon frequently observed in early pathogenesis \cite{kahn2000importance}. In this state, the network leverages intensified fluctuations to rescue pulsatility before a complete macroscopic collapse occurs.

Our demonstration of noise-induced macroscopic pacing reveals that biological networks do not merely tolerate intrinsic ion channel noise; they actively exploit it. In isolated cells, as shown in Section III.B, the required resonant noise severely disrupts the delicate intra-burst calcium spiking (spike rate), which is biologically detrimental to insulin exocytosis. However, by coupling through Cx36 gap junctions ($g_c$), the islet network effectively acts as a spatial low-pass filter \cite{sterk2024network, pourhosseinzadeh2025heterogeneity}. This network topology filters out erratic high-frequency spikes while preserving regular slow-wave bursts, ultimately restoring synchronized pulsatile insulin secretion in metabolically inhibited states.

To address the computational challenges in these systems, the introduction of the logarithmic centroid extraction method resolves the ``bathtub effect'' prevalent in highly flattened quasipotential landscapes. Traditional frameworks often struggle with local stochastic trapping in broad valleys, a phenomenon conceptually linked to challenges in anomalous diffusion \cite{metzler2000random} and manifold learning \cite{coifman2006diffusion}. By neutralizing zero-mean sampling noise in the logarithmic space, our approach provides a computationally efficient alternative to exhaustive global sampling. This framework offers potential extensions to broader non-equilibrium thermodynamic systems \cite{seifert2012stochastic, wang2008potential} and the detection of early warning signals for tipping points in rate-dependent models \cite{ashwin2012tipping, layritz2025early}.

Furthermore, to mitigate the functional impairment observed in isolated units, our network analysis demonstrates that gap-junction coupling acts as a vital structural filter. The sharp transition to macroscopic functional synchronization underscores how networks exploit localized fluctuations to overcome heterogeneity and deep metabolic inhibition. This aligns with recent evidence that structural connectivity and specific coupling schemes can significantly enhance coherence \cite{tessone2007theory, donofrio2019inhibition}. The current analytical framework provides a rigorous characterization of noise-induced dynamics within a basic two-cell homotypic motif. However, this fundamental model stands in contrast to actual physiological systems, such as intact pancreatic islets, where thousands of cells interact across intricate three-dimensional lattices or scale-free topologies. The ability of networks to leverage intrinsic noise for global coordination provides insights into synchronization phenomena across temporal multiplex hypernetworks \cite{PhysRevE.98.032305} and the optimization of multi-layer signal propagation \cite{charmpi2021optimizing}. The transition boundaries identified here suggest that applying varying coupling topologies to such large-scale architectures could give rise to diverse spatiotemporal patterns, including chimera states \cite{majhi2019chimera} or distinct permutation entropy signatures \cite{yan2019design}, which warrant further investigation. Furthermore, extending this analytical framework to more complex, heterogeneous islet models that incorporate additional intracellular metabolic pathways, such as mitochondrial dynamics, will be crucial to formally test the universality of these noise-induced rescue mechanisms.

\section{Conclusion}

\FloatBarrier

In this study, we systematically investigated the noise-induced dynamical transitions within a high-dimensional fast-slow pancreatic pacemaker model. By employing a semi-implicit integration scheme tailored for strictly bounded Feller diffusion, we circumvented numerical artifacts associated with unbounded stochastic simulations near physical limits. 

We proposed a logarithmic centroid extraction method to overcome the ``bathtub effect'' encountered when evaluating coherence resonance in deep energy valleys. This methodology successfully recovered the underlying adiabatic Kramers scaling with high precision ($R^2 > 0.95$) and explicitly identified the theoretical boundary where the strong-noise limit violates the adiabatic assumption. 

Finally, our extension to gap-junction coupled networks revealed a noise-induced phase transition from sub-threshold shivering to macroscopic functional synchronization. While this study establishes the qualitative mechanism of noise-induced synchronization via the competition between transverse contraction and stochastic diffusion, quantitatively mapping the strict stability boundaries remains an open challenge. Future investigations will focus on deriving quantitative analytical stability boundaries for the observed synchronization regimes and extending master stability function frameworks to bounded multiplicative noise, further bridging the gap between isolated single-cell Kramers scaling and collective network resilience. The quasipotential scaling laws and the centroid extraction framework established here offer a theoretical foundation for analyzing state failures in complex biological networks. Moving forward, these insights may inform the design of robust, noise-resilient architectures in broader multi-scale dissipative systems. Whether the spatial low-pass filtering and the logarithmic centroid extraction method can reveal higher-order topological phase transitions in such large-scale, spatially extended networks remains a compelling open question for future nonequilibrium statistical mechanics.

\begin{acknowledgments}
The author acknowledges the University of Sydney for providing the academic environment that inspired the initial ideas for this work.
\end{acknowledgments}

\section*{AUTHOR DECLARATIONS}

\subsection*{Conflict of Interest}

The author has no conflicts to disclose.

\subsection*{Author Contributions}

Yefan Wu: Conceptualization; Methodology; Software; Validation; Formal analysis; Investigation; Data curation; Visualization; Writing -- original draft; Writing -- review and editing.

\section*{DATA AVAILABILITY}

The data that support the findings of this study are available from the corresponding author upon reasonable request.

%aipnum4-2.bst 2019-01-14 (MD) hand-edited version of apsrev4-1.bst

\end{document}